# Origin of Charge Density Wave in Topological Semimetals SrAl$_4$ and EuAl$_4$

Lin-Lin Wang[1*], Niraj K. Nepal[1] and Paul C. Canfield[1,2]

[1]Ames National Laboratory, U.S. Department of Energy, Ames, IA 50011, USA
[2]Department of Physics and Astronomy, Iowa State University, Ames, IA 50011, USA

*llw@ameslab.gov

## Abstract

Topological semimetals in BaAl$_4$-type structure show many interesting behaviors, such as charge density wave (CDW) in SrAl$_4$ and EuAl$_4$, but not the isostructural and isovalent BaAl$_4$, SrGa$_4$ and BaGa$_4$. Here using Wannier functions based on density functional theory, we calculate the susceptibility functions with millions of *k*-points to reach the small *q*-vector and study the origin and driving force behind the CDW. Our comparative study reveals that the origin of the CDW in SrAl$_4$ and EuAl$_4$ is the strong electron-phonon coupling interaction for the transverse acoustic mode at small *q*-vector along the $\Gamma$-$Z$ direction besides the maximum of the real part of the susceptibility function from the nested Fermi surfaces of the Dirac-like bands, which explains well the absence of CDW in the other closely related compounds in a good agreement with experiment. We also connect the different CDW behaviors in the Al compounds to the macroscopic elastic properties.

# I. Introduction

The interplay between charge density wave (CDW) and topological phases has received much attention due to the perspective of coupling many-body interaction with non-trivial band structure topology. For example, the nesting Weyl points gapped by chiral symmetry breaking have been proposed to give arise to topological superconductivity[1, 2] or CDW with the latter to realize axion electrodynamics[3] for the case of opposite chirality or a topological phase with monopole harmonic order[4] for the same chirality. The Fermi surface nesting (FSN) among Weyl points has been recently studied for the CDW in the quasi-1D (TaSe$_4$)$_2$I[5, 6] and the long-wavelength helical magnetic order in NdAlSi[7]. Very recently CDW has been observed in the chiral CoSi[8, 9] with Kramer-Weyl points. CDW has also been used[10] to engineer topological phases to realize Dirac points close to the Fermi energy ($E_F$). However, besides FSN, the emergence of CDW also need the important ingredient of electron-phonon coupling[11, 12] (EPC). The traditional view of CDW comes from Peierls' instability[13] with perfect FSN in 1D to induce Kohn anomaly or soft phonon mode, which results in a structural transition at low temperature. This ideal 1D picture of CDW has been pervasive in condensed matter physics, but was recently substantially revised for real materials[14, 15]. Density functional theory[16, 17] (DFT) calculations[18, 19] and experiments[20, 21] on NbSe$_2$ and TaSe$_2$ have shown that the CDW in these compounds is determined by the real part of linear response or susceptibility function, Re$\chi(q)$, with a large EPC interaction to soften the acoustic phonon mode, while not by the FSN from the imaginary part of susceptibility function, Im$\chi(q)$, which gives a nesting vector not necessarily corresponding to the CDW $q$-vector in these specific compounds.

Here by studying the topological semimetals SrAl$_4$, EuAl$_4$ and the related compounds, we take a closer look at the microscopic mechanism for the CDW with topological non-trivial bands and find the critical role of EPC interaction or matrix element play for the CDW in SrAl$_4$ and EuAl$_4$ beyond just FSN of the nodal line Dirac-like bands. Topological semimetal compounds in the BaAl$_4$-type structure[22] have shown a range of interesting behaviors with a martensitic tetragonal-to-monoclinic structural transition in CaAl$_4$,[23] CDW in SrAl$_4$ and EuAl$_4$,[24-28] and multiple magnetic transitions, topological Hall effects and Skyrmions in EuAl$_4$ and related magnetic compounds at different temperature[25, 29-35]. The CDW in SrAl$_4$ and EuAl$_4$ has been measured to have a small $q$-vector along the $\Gamma$-$Z$ direction. But there is still a lack of understanding of the origin and driving force behind this CDW. Even more interesting is among the non-magnetic isostructural and isovalent series of SrAl$_4$, BaAl$_4$, SrGa$_4$ and BaGa$_4$, a recent experiment[24] has found that only SrAl$_4$ has CDW at 243 K. Such distinctly different behaviors regarding CDW make these topological semimetals the perfect subjects to test the proposed mechanism[14, 15] for CDW as being dominated by EPC interaction instead of FSN.

In this paper, we report a comparative computational study to reveal the origin of the CDW observed in SrAl$_4$ and EuAl$_4$, but its absence in BaAl$_4$ and the two Ga compounds. We calculate both Re$\chi(q)$ and Im$\chi(q)$ using Wannier functions with millions of $k$-points to reach the small $q$-vector. Although all of them are topological semimetals with nodal lines in the absence of spin-orbit coupling (SOC) and host Dirac points (DPs) with SOC as protected by the 4-fold rotational symmetry, these DPs are ~0.2 eV above the Fermi energy (E$_F$). The dominating features at E$_F$ instead are the Fermi surface (FS) shells formed by the tip and dip of the nodal line Dirac-like band dispersion with two nearby crossings of the E$_F$. With the imperfect FSN from $\chi(q)$, the FS shells give the maximum of the real part of susceptibility function, Re$\chi_{max}$, along the $\Gamma$-$Z$ direction with a small finite $q$-vector for all the three Al compounds, but not the Ga compounds. Then among SrAl$_4$, BaAl$_4$ and EuAl$_4$, the EPC calculations from density functional perturbation theory[36] (DFPT) show a larger EPC interaction or matrix element in SrAl$_4$ and EuAl$_4$ than BaAl$_4$ for the transverse acoustic (TA) mode at about the same $q$-vector of Re$\chi_{max}$ along the $\Gamma$-$Z$ direction to induce the TA mode to be imaginary. Our study reveals that the origin of the CDW in SrAl$_4$ and EuAl$_4$ is the strong EPC interaction for the TA mode at small $q$-vector along the $\Gamma$-$Z$ direction, besides the Re$\chi_{max}$ from the nested FS shells of nodal line Dirac-like bands, which explains well the absence of CDW in the other closely related isostructural and isovalent compounds, in a good agreement with experiment[24]. The different behaviors of the TA mode CDW can also be better understood in connection to the different macroscopic shear modulus and Poisson ratio. We find that the electron charge density redistribution between the Al network and different cation layers indicates the more ionic interaction in BaAl$_4$ than SrAl$_4$ and EuAl$_4$, thus can explain these different CDW and elastic behaviors. It is interesting to find the link between the TA mode CDW with microscopic EPC interaction and the macroscopic elastic properties.

## II. Results and Discussion

### II-a. Topological band structures

Using SrAl$_4$ as an example, the body-centered tetragonal (Pearson symbol tI10) crystal structure in space group 139 (*I4/mmm*) is shown in Fig.1(a). The bulk first Brillouin zone (BZ) and (001) surface BZ with the high symmetry *k*-points are displayed in Fig.1(b). The two Al sites are non-equivalent with 4*d* at (0.5, 0.0, 0.25) and 4*e* at (0.0, 0.0, *z*). These Al sites form a network of buckled square lattices sandwiched between the electron-donating Sr layers at the 1*a* site. Such structure is the binary version of the Fe-based superconductors and the related 122 compounds, where the two non-equivalent 4*d* and 4*e* sites are replaced by 3*d* transition metals (Fe, Co, and Ni) and main group V elements (As, P), respectively. Noticeably, the shortest Al-Al distance of 2.598 Å (DFT 2.616 Å) is between the two 4*e* sites from the two neighboring buckled square lattices along the 4-fold *c*-axis, which is slightly shorter than the Al-Al distance of 2.690 Å (DFT 2.688 Å) among the same buckled square lattice (see Supplementary Table 1).

The DFT-calculated band structure of SrAl$_4$ is plotted in Fig.1(c). The highest valence and lowest conduction bands according to simple filling are highlighted in blue and red, respectively. The multiple sawtooth Dirac-like dispersion of the valence and conduction bands near the E$_F$ point to a topological semimetal with nodal lines. Indeed, as plotted in Fig.1(e) for the band gap contours on the (110) plane without SOC, there are a nodal loop around the *Z* point and also nodal lines away from the *Γ-Z* direction toward the *X* point. More nodal lines and loops are plotted on (001) and (100) planes in Supplementary Figure 1. These nodal lines and Dirac-like dispersion are protected by the different sets of mirror symmetries, which are similar to the topological semimetal series of ZrSiS and the related compounds with square lattices of *p*-orbitals[37]. With SOC, these crossings in nodal lines are all gapped out except for the one along the *Γ-Z* direction as zoomed in Fig.1(d). The point group symmetry is D$_{4h}$. We have used Vasp2trace[38, 39] to analyze the elementary band representations. From the double group representation of D$_{4h}$ for the spinful system, the top valence band along the *Γ-Z* direction with the 4-fold rotation switches between two different 2-dimensioinal irreducible representations of Λ$_7$ and Λ$_6$, thus there is no mixing guarantied at the band crossing, or equivalently speaking, the pair of DPs is protected by the 4-fold rotational symmetry[40-45]. Our results also agree with the previous study[22] on BaAl$_4$ for the DPs along the $\Gamma$-$Z$ direction.

The locations of the DPs are also indicated in the BZ by the red dots in Fig.1(b) at the momentum energy of (0, 0, ±0.1912 Å$^{-1}$; $E_F$+0.2186 eV). When projected on the (001) surface, the pair of DPs overlap, but there are still topological surface states converging to the DP projection as shown in Fig.1(f) along the $\bar{\Gamma}$-$\bar{X}$ direction. The spin texture in Fig.1(g) shows that these surface states are spin-momentum locked as expected for topological non-trivial surface states. Furthermore, we have also plotted these topological surface states in the other direction of $\bar{\Gamma}$-$\bar{M}$ in Fig.1(h) with quite different band dispersion, but still they converge to the same DP projection. Because the DPs are 0.22 eV above the $E_F$, the physical properties such as CDW is not directly determined by the DPs, but rather the 3D FS of the nodal line Dirac-like band dispersion near the $E_F$.

## II-b. Susceptibility functions and phonon dispersions

The 3D FS of SrAl$_4$ is plotted in Fig.2(a). The valence band FS with hole pockets (yellow outside and blue inside) are centered around the $Z$ point, while the conduction band FS with electron pockets (purple outside and green inside) are around the $\Gamma$ point. Because of the Dirac-like dispersion, when the bands cross the $E_F$ twice nearby along the same high symmetry directions with a dip or tip of the Dirac-like bands, they form FS of thin shells instead of space-filled objects. Such features can be clearly seen when the 3D FS are separated into the hole and electron pieces with a near top-view in Fig.2(b) and (c), respectively. To study FSN and CDW from band structure, the bare susceptibility function[14], $\chi(\boldsymbol{q})$, based on DFT single-particle Kohn-Sham bands needs to be calculated in real and imaginary parts,

$$Re\chi(\boldsymbol{q}) = \sum_{n,m,\boldsymbol{k}} \frac{f(\varepsilon_{n,\boldsymbol{k}}) - f(\varepsilon_{m,\boldsymbol{k}+\boldsymbol{q}})}{\varepsilon_{n,\boldsymbol{k}} - \varepsilon_{m,\boldsymbol{k}+\boldsymbol{q}}} \qquad (1)$$

$$\lim_{\omega \to 0} Im\chi(\boldsymbol{q},\omega)/\omega = \sum_{n,m,\boldsymbol{k}} \delta(\varepsilon_{n,\boldsymbol{k}})\delta(\varepsilon_{m,\boldsymbol{k}+\boldsymbol{q}}) \qquad (2)$$

where $f(\varepsilon_{n,\boldsymbol{k}})$ and $\delta(\varepsilon_{n,\boldsymbol{k}})$ are the Fermi-Dirac distribution and Delta functions, respectively, of $\varepsilon_{n,\boldsymbol{k}}$ the $n$-th band energy eigenvalue at the $\boldsymbol{k}$ point with the $E_F$ set to zero. The calculation of $\chi(\boldsymbol{q})$ requires very dense double meshes ($\boldsymbol{k}$ and $\boldsymbol{q}$). Using the maximally localized Wannier functions[46, 47] (MLWF), we are able to calculate the 3D susceptibility function $\chi(q)$ efficiently with millions of $k$-points in the BZ.

The real part of the 3D susceptibility function $\text{Re}\chi(q)/\text{Re}\chi_{max}$ of SrAl$_4$ is shown in Fig.2(d) for the whole BZ with its maximum appearing as two small red spheres along the $\Gamma$-$Z$ direction with a small finite $q$-vector. As plotted along the $\Gamma$-$Z$ direction in 1D in Fig.2(g), the calculated $q$-vector for $\text{Re}\chi_{max}$ is 0.23($\pi/c$) for SrAl$_4$, agreeing well with the experimental data of 0.22($\pi/c$) (dashed line) for the observed incommensurate CDW at 243 K[24]. Such a CDW $q$-vector has usually been associated with FSN, i.e., the sharp peaks in the imaginary part of susceptibility function, $\text{Im}\chi(q)$, especially for 1D cases. But as plotted in Fig.2(e) along the $\Gamma$-$Z$ direction, the peak in the $\text{Im}\chi(q)$ of SrAl$_4$ around 0.23($\pi/c$) is not sharp, indicating an imperfect nesting due to the 3D nature of the FS in real materials. The imperfect nesting vector along the $k_z$ direction in SrAl$_4$ arises from the pyramid-shape valence FS shell in Fig.2(b) and is also indicated by an orange arrow in the 2D FS contour on the (110) plane in Fig.2(f). The calculated phonon band dispersion of SrAl$_4$ is plotted in Fig.2(h) showing an imaginary TA phonon mode (black arrow) along the $\Gamma$-$Z$ direction with a $q$-vector of 0.24($\pi/c$), which matches that of calculated $\text{Re}\chi_{max}$. This match of the $q$-vector of the $\text{Re}\chi_{max}$ and imaginary TA mode in SrAl$_4$ shows that $\text{Re}\chi_{max}$ and the imperfect FSN can be an indicator for CDW. However, we will later show that imperfect FSN and the $\text{Re}\chi_{max}$ does not necessarily guarantee a CDW in the case of BaAl$_4$, because BaAl$_4$ lacks a strong enough EPC to induce a soft phonon mode.

For the other three non-magnetic isostructural and isovalent compounds of BaAl$_4$, SrGa$_4$ and BaGa$_4$, their band structures and FS are plotted in Fig.3 for comparison to SrAl$_4$. BaAl$_4$ has a very similar band structure and FS to SrAl$_4$, as shown in Fig.3(a) and (d), only with small difference at the BZ boundaries. BaAl$_4$ retains the pyramid-shape valence FS shell around the $Z$ point and the interlocked pattern of the hole and electron FS shells. The $\text{Im}\chi(q)$ of BaAl$_4$ as plotted in Fig.2(e), also has a maximum along the $\Gamma$-$Z$ direction, which has a larger $q$-vector and with a more extended plateau than that of SrAl$_4$. In contrast, the $\text{Re}\chi_{max}$ of BaAl$_4$ in Fig.2(g) is at a smaller $q$-vector than that of SrAl$_4$. This shows a more imperfect FS nesting in BaAl$_4$ than SrAl$_4$ and the mismatch of the peaks in $\text{Im}\chi(q)$ and $\text{Re}\chi(q)$ is not unexpected, because the latter include contributions from the bands away from the $E_F$. As plotted in Supplementary Figure 2, the phonon band dispersion of BaAl$_4$ does not have an imaginary phonon mode or CDW near the Re$\chi_{max}$ $q$-vector.

Next for SrGa$_4$ in Fig.3(b) and (e), besides the extra band inversion at the $S$ point, the main difference is the upshift of the valence bands along the $\Gamma$-$Z$ direction. The inner shell of the hole pockets at the $Z$ point almost merge into a single large pocket with the next valence band touching the $E_F$ along the $Z$-$S_1$ direction to form a small ring-shape pocket (red) around the $Z$ point. As the results, the Re$\chi_{max}$ $q$-vector along the $\Gamma$-$Z$ direction for SrGa$_4$ is reduced toward $q=0$ as plotted in Fig.2(g). SrGa$_4$ does not have a CDW either as shown by the absence of imaginary phonon dispersion in Supplementary Figure 2. Moving to BaGa$_4$ in Fig.3(c), the top valence band along the $\Gamma$-$Z$ direction is pushed to even higher energy with the next valence band now crossing the $E_F$ to give a squarish hole pocket (red) centered at the $Z$ point as seen in Fig.3(f). The imperfect FSN along the $\Gamma$-$Z$ direction now totally disappears, resulting in no maximum of Re$\chi(q)$ near the $\Gamma$ point for BaGa$_4$ in Fig.2(g). Although the Re$\chi_{max}$ for BaGa$_4$ is now at the $Z$ point, there are no imaginary phonon modes as shown in Supplementary Figure 2, thus no CDW either. These results from the calculated Re$\chi(q)$ and phonon dispersion show only CDW in SrAl$_4$ but not the other three non-magnetic compounds with the same crystal structure and number of valence electrons, which agree well with experimental observations[24].

## II-c. Electron-phonon coupling and CDW

Despite the similar band structure, FS and also the finite small $q$-vector for the Re$\chi_{max}$ in BaAl$_4$ when compared to SrAl$_4$, the absence of CDW in BaAl$_4$ demands a closer look. To reveal the origin of the observed CDW in SrAl$_4$ but not BaAl$_4$, using density functional perturbative theory (DFPT), we have calculated the phonon lifetime $\gamma_{qv}$ due to EPC, i.e. the imaginary part of the phonon self-energy, which is proportional to both the EPC matrix elements $|g(q)|^2$ and Im$\chi(q)$. As plotted in Fig.4(a) with different electronic smearing, the $\gamma_{qv}$ for the TA mode along the $\Gamma$-$Z$ direction in SrAl$_4$ is two to three times larger than that in BaAl$_4$. To put the EPC of the TA mode into perspective among all the other modes, $\gamma_{qv}$ can be converted into the dimensionless mode-resolved EPC strength as in $\lambda_{qv} = \gamma_{qv}/(\pi N(E_F)\omega_{qv}^2)$, where $N(E_F)$ is the electronic density of states at $E_F$. With the electronic smearing of 0.04 Ry to stabilize the imaginary TA mode in SrAl$_4$, the calculated $\lambda_{qv}$ for both SrAl$_4$ and BaAl$_4$ are plotted on top of their phonon dispersion in Fig.4(c) and (d) for comparison. As seen from the green shade for $\lambda_{qv}$, the largest difference is that SrAl$_4$ (Fig.4(c)) has a much larger $\lambda_{qv}$ than BaAl$_4$ (Fig.4(d)) for the TA mode. Unlike SrAl$_4$, the $\lambda_{qv}$ of BaAl$_4$ for the TA mode only has similar magnitude to the other modes near the $\Gamma$ and $Z$ points. The eigenvectors of the TA mode (E$_u$) and three such zone-center optical modes E$_g$ (2.3 THz), B$_{1g}$ (6.9 THz) and A$_{1g}$ (10.8 THz) of SrAl$_4$ are shown in Fig.4(e). The E$_g$ mode corresponds to the in-plane motion of Al network, while the B$_{1g}$ and A$_{1g}$ modes at higher frequency correspond to out-of-plane motion of Al network. EPC interaction also directly affects the size of the phonon softening via the real part of phonon self-energy. The $\lambda_{qv}$ decorated plots show that SrAl$_4$ is subjected to a larger size of phonon softening than BaAl$_4$. Given the small frequency of the TA mode near $\Gamma$, the larger phonon softening in SrAl$_4$ can induce this mode to be imaginary and give the CDW instability. As plotted in Fig.4(b) for the dispersion of TA mode with different electronic smearing, in contrast to BaAl$_4$, this mode in SrAl$_4$ has a substantial softening with the decreasing smearing size (mimicking the electronic temperature) due to the stronger EPC interaction. At the smearing of 0.02 Ry, the TA mode becomes imaginary at a $q$-vector around 0.24($\pi/c$), in a good agreement with the experimentally observed $q_{CDW}$[24].

When Sr is replaced by Ba going from SrAl$_4$ to BaAl$_4$ and with the similar FSN, the conventional wisdom is that with a larger mass of Ba than Sr giving lower phonon frequencies, the phonon softening to imaginary mode should be easier and CDW should also exist in BaAl$_4$, assuming EPC interaction stay almost the same. However, both these assumptions are not true for the TA mode along the $\Gamma$-$Z$ direction here. Firstly, comparing the phonon dispersion of SrAl$_4$ (Fig.4(c)) to BaAl$_4$ (Fig.4(d)), the highest phonon energy is reduced from 10.8 THz in SrAl$_4$ to 9.9 THz in BaAl$_4$ and the overall phonon energy range is rescaled by the mass factor as expected. Along the $\Gamma$-$Z$ direction, the longitudinal acoustic (LA) mode is also reduced from 3.0 THz in SrAl$_4$ to 2.3 THz in BaAl$_4$ at the $Z$ point. However, the TA mode changes in the opposite direction and increases from 1.4 THz in SrAl$_4$ to 1.8 THz in BaAl$_4$ at the $Z$ point as also shown in Fig.4(b). The first optical E$_g$ mode at $\Gamma$ for the in-plane motion of Al network also increases from 2.3 THz in SrAl$_4$ to 2.7 THz in BaAl$_4$. Secondly, comparing the TA mode lifetime $\gamma_{qv}$ between BaAl$_4$ and SrAl$_4$ in Fig.4(a), the lifetime of the TA mode in BaAl$_4$ becomes smaller than that in SrAl$_4$ for both large and small $q$-vectors. Note the lifetime $\gamma_{qv}$ accounting for the EPC interaction has no mass dependence through $\omega_{qv}$ by definition. It evaluates the deformation of potential experienced by the electrons on FS due to the structural distortion from phonon modes as convoluted by the Im$\chi(q)$. The same EPC matrix elements of the deformed potential are also used to calculate the real part of the Green's function for the phonon softening via convolution with Re$\chi(q)$. In Fig.4(a), lifetime $\gamma_{qv}$ shows that the EPC interaction is two to three times smaller in BaAl$_4$ than SrAl$_4$ for the TA mode along the $\Gamma$-$Z$ direction. Thus, when going from SrAl$_4$ to BaAl$_4$, both the increase in the single-particle (bare) phonon frequency and the decrease in EPC interaction of the TA mode work against the phonon softening in BaAl$_4$. As the results, unlike SrAl$_4$, there is no CDW in BaAl$_4$.

The above analysis focusing on EPC interaction besides imperfect FSN can also explain the behavior of EuAl$_4$, which has a CDW at ~140 K above the Neel temperature, albeit with an even larger mass of Eu than Ba. The band structure and FS of the non-magnetic EuAl$_4$ (no Eu 4$f$) in Fig.5(a) and (b) resemble those of SrAl$_4$ in Fig.1(c) and 2(a), respectively. The DP has the momentum-energy at (0, 0, ±0.1928 Å$^{-1}$; $E_F$+0.1508 eV). The FS shell of the EuAl$_4$ conduction band is similar to that of SrAl$_4$, except for the larger electron pocket at the $N$ point than SrAl$_4$. The FS shell of the EuAl$_4$ valence band is also similar to that of SrAl$_4$, giving a similar imperfect FSN in $\chi(q)$ as plotted in Fig.5(d) and (e) with a slightly smaller nesting $q$-vector at 0.19($\pi$/c) than SrAl$_4$. Noticeably, the Re$\chi_{max}$ of EuAl$_4$ has a relatively narrow peak similar to SrAl$_4$ at the small q-vector, rather than the extended plateau of BaAl$_4$. The 3D Re$\chi(q)$/Re$\chi_{max}$ of EuAl$_4$ plotted in the full first BZ in Fig.5(c) shows the maximum is indeed only along the $\Gamma$-$Z$ direction. For the TA mode dispersion plotted in Fig.5(e) in comparison to SrAl$_4$ and BaAl$_4$, although in term of atomic mass Sr < Ba < Eu, the TA mode of BaAl$_4$ is the highest among the three in the order of SrAl$_4$ ~ EuAl$_4$ < BaAl$_4$ at the $Z$ point. Then although they all have similar Re$\chi_{max}$ $q$-vector along the $\Gamma$-$Z$ direction in Fig.5(f), the three EPC interactions of the TA mode are quite different. As plotted in Fig.5(g), the $\gamma_{qv}$ of EuAl$_4$ has an interesting behavior as a function of $q$ when compared to SrAl$_4$ and BaAl$_4$. At small-$q$ (<0.2), the $\gamma_{qv}$ of EuAl$_4$ increases fast like SrAl$_4$, but then plateau like BaAl$_4$ and next decreases at large-$q$ to be even smaller than BaAl$_4$. Although the maximum of $\gamma_{qv}$ for the TA mode in EuAl$_4$ is smaller than that of BaAl$_4$, its value at small-$q$ (<0.2) instead is larger than BaAl$_4$ and closer to that of SrAl$_4$. The small $q$-vector range of the EPC interaction is the most relevant to CDW because of the small Re$\chi_{max}$ $q$-vector. Thus, the EPC strength shown by $\gamma_{qv}$ at small-$q$ (<0.2) is in the order of SrAl$_4$ > EuAl$_4$ > BaAl$_4$, explaining why both SrAl$_4$ and EuAl$_4$ have CDW, while BaAl$_4$ does not, even though all three have imperfect FSN and Re$\chi_{max}$ $q$-vector.

The TA mode CDW here involves a local shear distortion perpendicular to the $c$-axis. It is very interesting to notice that among the calculated bulk elastic properties (see Table.1), the bulk modulus (B) of 50.6 GPa for BaAl$_4$ is slightly smaller than the 52.9 GPa for SrAl$_4$, reflecting a larger cation size of Ba than Sr, giving both larger $a$ and $c$ lattice constants with a slightly smaller $c/a$ ratio. However, the shear modulus (G) of 36.8 GPa for BaAl$_4$ is larger than the 29.1 GPa for SrAl$_4$. This corresponds to a much smaller Poisson ratio of 0.207 for BaAl$_4$ than the 0.268 for SrAl$_4$, which means a compression along the $c$-axis has a less in-plane expansion in response for BaAl$_4$ than SrAl$_4$, i.e., the in-plane interaction is stiffer for BaAl$_4$ than SrAl$_4$. Then moving to EuAl$_4$ with the smallest lattice constants among the three, both B of 57.1 GPa and G of 33.9 GPa increase comparing to SrAl$_4$. But the G of EuAl$_4$ is still smaller than that of BaAl$_4$, resulting in a Poisson ratio of 0.252 similar to that of SrAl$_4$, not BaAl$_4$. Thus, the in-plane interaction in EuAl$_4$ is still softer than BaAl$_4$.

To better understand these differences from electronic structure, we have plotted and compared the electron charge density difference of $\rho(XAl_4) - \rho(X) - \rho(Al_4)$ for X=Sr, Ba and Eu, respectively in Fig.5 (h-j). The charge density redistributions between the Al network and the different cation layers show that there is more electron transferred from the Ba layer to Al network and also more charge accumulation (yellow) at the boundary between the Ba and Al network than the cases of Sr and Eu. The more ionic character of the interaction in BaAl$_4$ with more in-plane charge accumulation makes it harder for the in-plane shear distortion between the Al network and Ba layer, which explains a much smaller Poisson ratio for BaAl$_4$ than SrAl$_4$ and EuAl$_4$. This also means the TA mode softening for CDW with the local shear distortion in BaAl$_4$ is much harder than that in SrAl$_4$ and EuAl$_4$. It is interesting to find the connection between the CDW with microscopic EPC interaction and the macroscopic elastic properties.

## III. Conclusions

In conclusion, using the approaches based on density functional theory, we have explained the origin of charge density wave (CDW) in SrAl$_4$ and EuAl$_4$ in comparison to the BaAl$_4$, SrGa$_4$ and BaGa$_4$. Although all of them have nodal lines in the absence of spin-orbit coupling (SOC) and become Dirac semimetals with SOC, the Dirac points are ~0.2 eV above the Fermi energy (E$_F$). The Dirac-like dispersion of the valence and conduction bands form Fermi surface (FS) shells providing imperfect FS nesting (FSN) along the $\Gamma$-$Z$ direction for the three Al compounds, but not the Ga compounds due to the second valence band being pushed up to cross the E$_F$. The susceptibility functions, $\chi(q)$, have been calculated efficiently on a dense mesh of millions of *k*-points with the maximally localized Wannier functions (MLWF). The imperfect FSN is verified by the maximum of the real part of susceptibility function, Re$\chi_{max}$, along the $\Gamma$-$Z$ direction with a small *q*-vector for the Al compounds, while it is absence in the Ga ones. Then among the three Al compounds, the electron-phonon coupling (EPC) calculations show a large EPC interaction at small *q*-vector for the transverse acoustic (TA) mode along the $\Gamma$-$Z$ direction for SrAl$_4$ and EuAl$_4$, which provides the driving force for the CDW to soften the TA mode, while it is absence in BaAl$_4$. Our study reveals that the origin of the CDW in SrAl$_4$ and EuAl$_4$ is the strong EPC interaction for the TA mode along the $\Gamma$-$Z$ direction at small *q*-vector, besides the Re$\chi_{max}$ from the nested FS shells of nodal line Dirac-like bands, which explains well the experimental observations. We also connect the different TA mode CDW distortion to the different macroscopic shear modulus and Poisson ratio. We find that the electron charge density redistribution between the Al network and different cation layers indicates the more ionic interaction in BaAl$_4$ than SrAl$_4$ and EuAl$_4$, thus explains these different CDW and elastic behaviors.

## IV. Methods

Density functional theory[16, 17] (DFT) calculations have been performed with PBE[48] exchange-correlation functional including SOC using a plane-wave basis set and projector augmented wave method[49], as implemented in the Vienna Ab-initio Simulation Package[50, 51] (VASP). We use a kinetic energy cutoff of 300 eV, $\Gamma$-centered Monkhorst-Pack[52] (11×11×11) k-mesh, and a Gaussian smearing of 0.05 eV. The ionic positions and unit cell vectors are fully relaxed with the remaining absolute force on each atom being less than $1\times10^{-4}$ eV/Å. The DFT-calculated lattice parameters with PBE+SOC are within ~ 2% of the experimental data, showing good agreement (see Supplementary Table 1). Phonon band dispersions have been calculated with finite difference method using PHONOPY[53] on a (3×3×4) supercell of the conventional cell with 360 atoms on the k-mesh of (4×4×2). To calculate susceptibility functions efficiently, maximally localized Wannier functions (MLWF)[46,47] and the tight-binding model have been constructed to reproduce closely the band structure within ±1eV of the Fermi level by using Group II *sd* and Group III *sp* orbitals. The susceptibility functions in Eqn.(1) and (2) have been calculated with four bands, two valence and two conduction bands, around the $E_F$ on the dense (120×120×90) k and q-mesh using the MLWFs, where the Fermi-Dirac distribution is sampled at the temperature of 100 K and the Delta functions are approximated with Gaussian functions with a smearing of 0.02 eV. The surface spectral functions have been calculated with the surface Green's function methods[54,55] as implemented in WannierTools[56]. Electron-phonon coupling (EPC) interaction has been calculated with density functional perturbative theory (DFPT) as implemented in Quantum Espresso[57] (QE) using ultra-soft pseudopotentials with a kinetic energy cutoff of 50 Ry, a (3×3×3) q-mesh and (9×9×9) k-mesh with different smearing. In Supplementary Figure 3 we show the convergence of calculated phonon band dispersions of $SrAl_4$ with respect to increased k-mesh to (18×18×18) and q-mesh to (4×4×4) in QE using DFPT and also a larger supercell of (4×4×4) of the conventional cell with 640 atoms and increased k-mesh in VASP using the finite displacement method with PHONOPY. The second-order elastic properties have been calculated using the stress-strain methodology[58,59]. The crystal structures and electron charge density difference have been visualized with VESTA[60].

**Data Availability**: The data that support the findings of this study are available from the corresponding author upon reasonable request.

## Acknowledgements

The susceptibility function calculations in this work at Ames National Laboratory were supported by the U.S. Department of Energy, Office of Science, Basic Energy Sciences, Materials Sciences and Engineering Division. Topological band structure analysis was supported by the Center for the Advancement of Topological Semimetals, an Energy Frontier Research Center funded by the U.S. Department of Energy Office of Science, Office of Basic Energy Sciences through the Ames National Laboratory under its Contract No. DE-AC02-07CH11358. Electron-phonon coupling calculations were supported by the Ames National Laboratory LDRD. The Ames National Laboratory is operated for the U.S. Department of Energy by Iowa State University under Contract No. DE-AC02-07CH11358.

**Author Contributions**: P.C.C. and L.-L.W. conceived and designed the work. L.-L.W. designed and performed the ab initio calculations on susceptibility functions and topological band structure analysis. L.-L.W. and N.K.N. performed electron-phonon coupling and elastic calculations. All authors discussed the results and contributed to the final manuscript.

**Competing Interests**: The authors declare no competing interests.

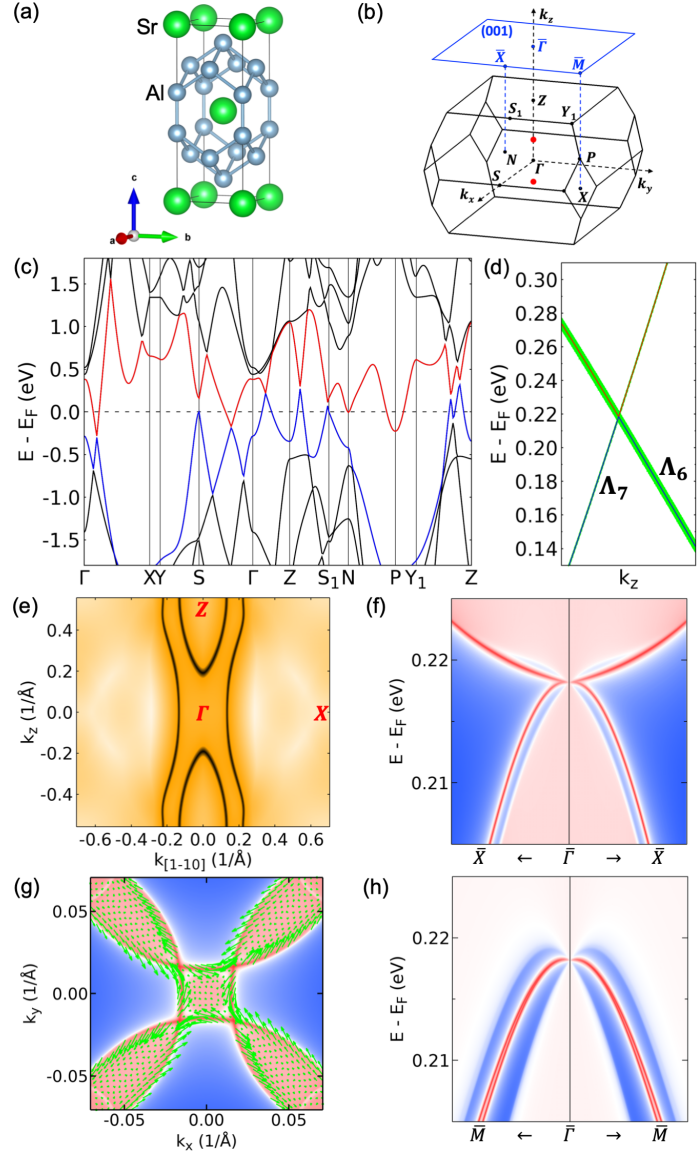

Figure 1. Topological electronic band structures of SrAl$_4$. (a) Crystal structure of SrAl$_4$ in the body-centered tetragonal (tI10) of space group 139 (*I4/mmm*). Sr (Al) is in large green (small blue) spheres. (b) Bulk Brillouin zone (BZ) with high symmetry *k*-points and those on (001) surface BZ are labeled. A pair of Dirac points (DPs) along $\Gamma$-$Z$ direction are indicated by red dots. (c) SrAl$_4$ band structure calculated in density functional theory (DFT) with spin-orbit coupling (SOC). The highest valence and lowest conduction bands are in blue and red, respectively. The dispersion along $\Gamma$-$Z$ is zoomed around the DP in (d) with the irreducible representations labeled and also the Al $p_z$ orbital projection at 4*e* site in green shade. (e) Nodal lines on the (110) plane shown by the zero gap. (f) Surface spectral function along the (001) $\bar{\Gamma}$-$\bar{X}$ direction. (g) (001) 2D Fermi surface at $E_F$+0.21 eV with spin texture shown in green arrows. (h) Surface spectral function along the (001) $\bar{\Gamma}$-$\bar{M}$ direction showing the topological surface states converging to the projection of DPs in different directions together with (f).

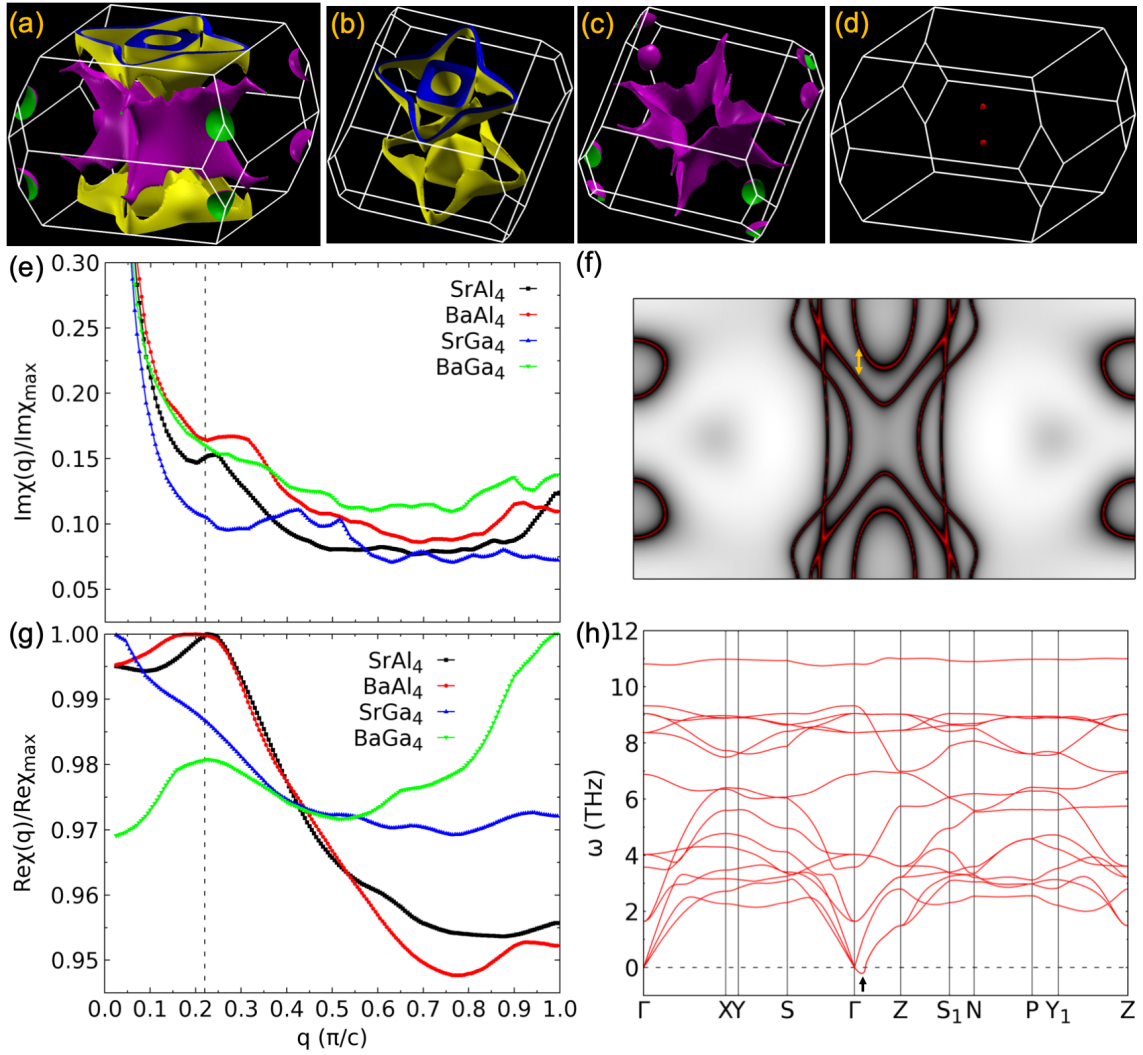

Figure 2. Fermi surface, susceptibility functions, phonon dispersion and charge density wave of SrAl$_4$. (a) SrAl$_4$ Fermi surface (FS) consists of valence (yellow outside and blue inside) and conduction band (purple outside and green inside) from the side-view. The individual FS pieces of valence and conduction bands from the near top-view are in (b) and (c), respectively. (d) 3D real part of the susceptibility function ($\chi(q)$) as in Re$\chi(q)$/Re$\chi_{max}$ of SrAl$_4$ with the iso-surface value of 0.999 showing the maximum at $q=\pm 0.23(\pi/c)$. (e) The imaginary part of the susceptibility functions Im$\chi(q)$/Im$\chi_{max}$ along the $\Gamma$-$Z$ direction for the four compounds. The vertical dashed line marks the experimental charge density wave (CDW) vector for SrAl$_4$ at $q_{CDW}=0.22(\pi/c)$. (f) 2D FS cut on the (110) plane showing the FS shells of the valence and conduction bands in SrAl$_4$ with the nesting vector marked by an orange arrow. The high (low) intensity is in red (grey). (g) The real part of the susceptibility functions Re$\chi(q)$/Re$\chi_{max}$ along the $\Gamma$-$Z$ direction. (h) Phonon band dispersion $\omega(q)$ of SrAl$_4$ with imaginary transverse acoustic (TA) mode along the $\Gamma$-$Z$ direction (black arrow).

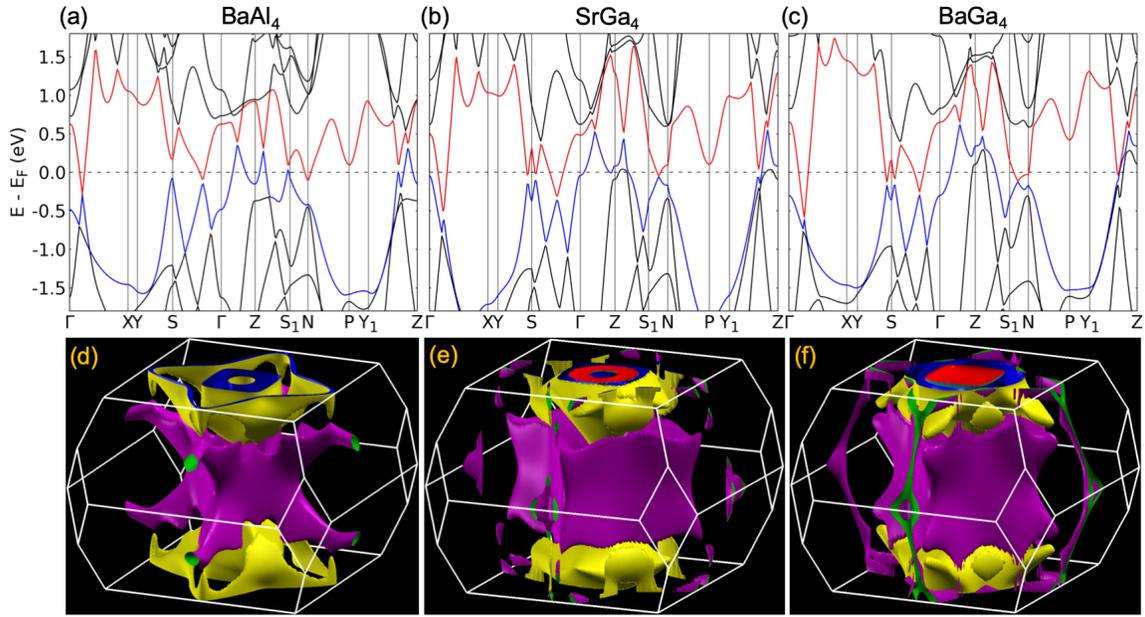

Figure 3. Electronic band structure and Fermi surface of the other three alkali earth compounds. For the electronic band structure of (a) $BaAl_4$, (b) $SrGa_4$ and (c) $BaGa_4$, the highest valence and lowest conduction bands are in blue and red, respectively. For the 3D Fermi surface (FS) of (d) $BaAl_4$, (e) $SrGa_4$ and (f) $BaGa_4$, it consists of valence (yellow outside and blue inside) and conduction bands (purple outside and green inside) from the side-view.

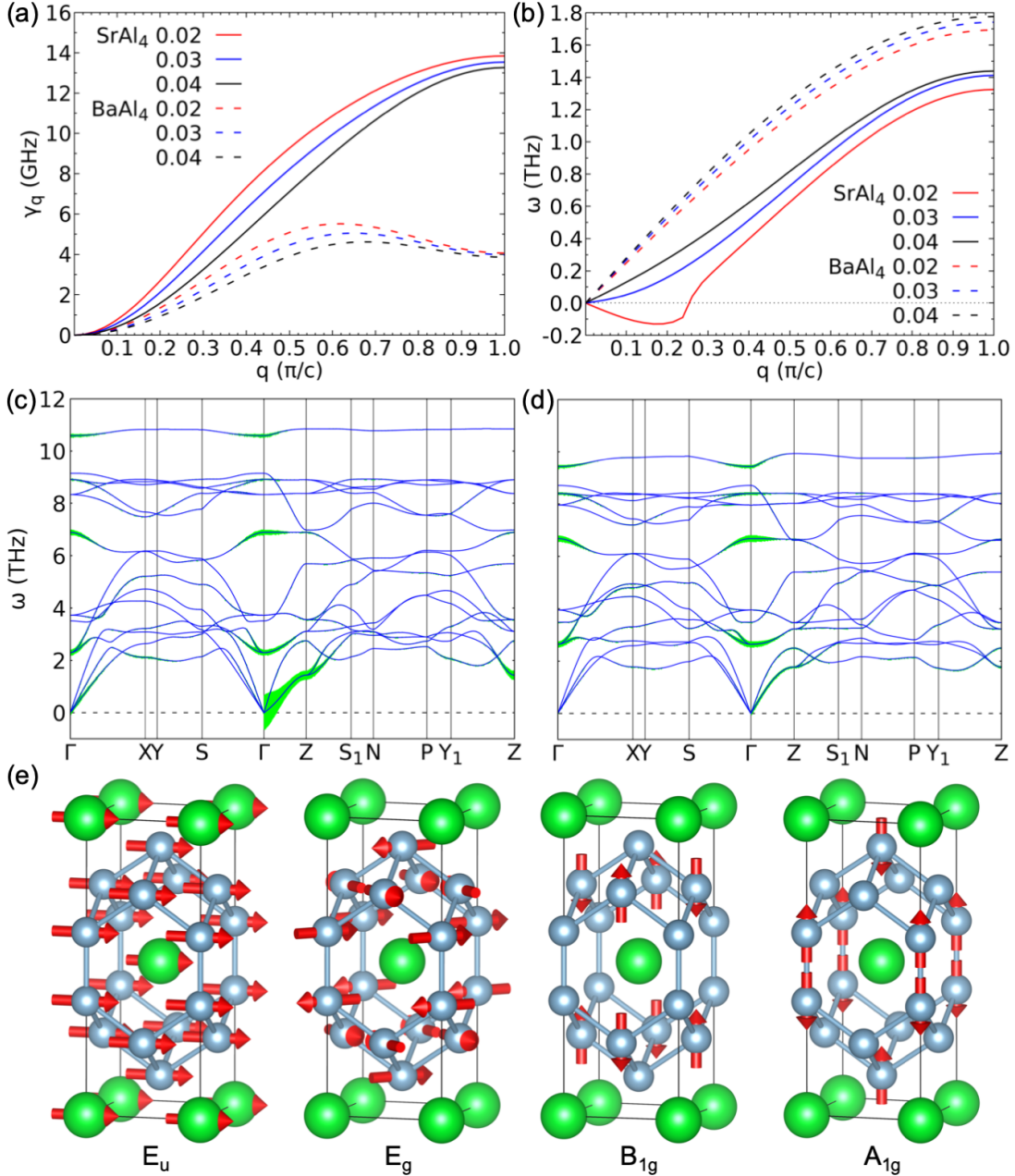

Figure 4. Comparison of electron-phonon coupling between SrAl$_4$ and BaAl$_4$. Calculated phonon (a) lifetime $\gamma_q$ and (b) dispersion $\omega$ for the transverse acoustic mode along the $\Gamma$-$Z$ direction with different smearing in Ry for SrAl$_4$ (solid lines) and BaAl$_4$ (dashed lines). (c) and (d) Phonon band dispersion decorated with the mode-resolved electron-phonon coupling (EPC) strength ($\lambda_q$) in green shadow for SrAl$_4$ and BaAl$_4$, respectively, with the electronic smearing of 0.04 Ry. (e) Eigenvectors of the four zone-centered phonon modes with sizable $\lambda_q$ from low to high frequency as shown in (c) and (d). The atomic displacements are indicated by red arrows and the corresponding irreducible representations of E$_u$, E$_g$, B$_{1g}$ and A$_{1g}$ are labeled.

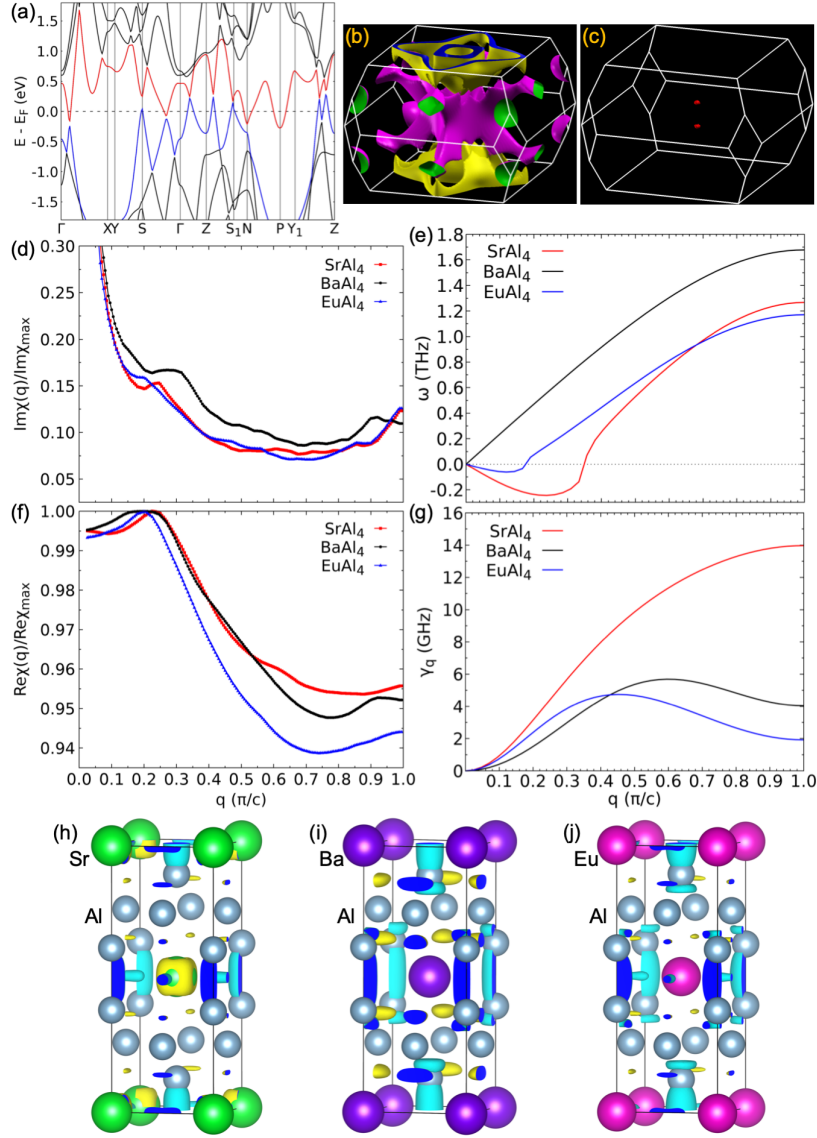

Figure 5. Band structure, electron-phonon coupling and bonding in EuAl$_4$ with comparison to SrAl$_4$ and BaAl$_4$. (a) EuAl$_4$ bulk band structure with the top valence and conduction bands plotted in blue and red, respectively. (b) Fermi surface (FS) consists of valence (yellow outside and blue inside) and conduction bands (purple outside and green inside) from the side-view. (c) 3D real part of susceptibility function $\text{Re}\chi/\text{Re}\chi_{max}$ with the maximum value shown as red pockets. (d) The 1D imaginary part of susceptibility function $\text{Im}\chi(q)/\text{Re}\chi_{max}$ plotted along the $\Gamma$-$Z$ direction ($q$) for SrAl$_4$, BaAl$_4$ and EuAl$_4$. (e) Calculated phonon dispersion $\omega$ of the transverse acoustic (TA) mode along the $\Gamma$-$Z$ direction with an electronic smearing of 0.015 Ry for the three compounds. (f) The 1D real part of susceptibility function $\text{Re}\chi(q)/\text{Re}\chi_{max}$. (g) The lifetime $\gamma_q$ for the TA mode. (h-j) Electron charge density of $\rho(XAl_4) - \rho(X) - \rho(Al_4)$ for X=Sr, Ba and Eu, respectively. The positive (negative) iso-surfaces of $2.1\times10^{-3}$ (e/Å$^3$) are shown in yellow (cyan) on the outside and blue from inside.

|  | SrAl$_4$ | BaAl$_4$ | EuAl$_4$ |
|---|---|---|---|
| Bulk modulus (B) (GPa) | 52.9 | 50.6 | 57.1 |
| Shear modulus (G) (GPa) | 29.1 | 36.8 | 33.9 |
| Poisson ratio ($\nu$) | 0.268 | 0.207 | 0.252 |

Table 1. Calculated elastic properties of SrAl$_4$, BaAl$_4$ and EuAl$_4$ in density functional theory (DFT).

# Supplementary Information

# Origin of Charge Density Wave in Topological Semimetals SrAl$_4$ and EuAl$_4$

Lin-Lin Wang[1*], Niraj K. Nepal[1] and Paul C. Canfield[1,2]

[1]Ames National Laboratory, U.S. Department of Energy, Ames, IA 50011, USA
[2]Department of Physics and Astronomy, Iowa State University, Ames, IA 50011, USA

*llw@ameslab.gov

|  |  | *a* (Å) | *c* (Å) | *z* | 4*e*-4*e* (Å) | 4*d*-4*e* (Å) |
|---|---|---|---|---|---|---|
| SrAl$_4$ | Expt[1] | 4.461 | 11.209 | 0.3841 | 2.598 | 2.690 |
|  | PBE+SOC | 4.456 | 11.250 | 0.3838 | 2.616 | 2.688 |
|  | δ (%) | −0.11 | +0.37 | −0.08 | +0.69 | −0.07 |
| BaAl$_4$ | Expt[2] | 4.566 | 11.250 | 0.3800 | 2.700 | 2.711 |
|  | PBE+SOC | 4.562 | 11.326 | 0.3812 | 2.691 | 2.722 |
|  | δ (%) | −0.09 | +0.68 | +0.32 | −0.33 | +0.41 |
| EuAl$_4$ | Expt[3] | 4.402 | 11.163 | 0.380 | 2.679 | 2.636 |
|  | PBE+SOC | 4.380 | 11.186 | 0.3853 | 2.565 | 2.662 |
|  | δ (%) | −0.50 | +0.21 | +1.39 | −4.26 | +0.99 |
| SrGa$_4$ | Expt[4] | 4.4474 | 10.7300 | 0.38299 | 2.511 | 2.642 |
|  | PBE+SOC | 4.502 | 10.807 | 0.3822 | 2.546 | 2.666 |
|  | δ (%) | +1.22 | +0.72 | −0.21 | +1.39 | +0.91 |
| BaGa$_4$ | Expt[5] | 4.5661 | 10.7780 | 0.3799 | 2.589 | 2.678 |
|  | PBE+SOC | 4.621 | 10.835 | 0.3788 | 2.627 | 2.699 |
|  | δ (%) | +1.20 | +0.53 | −0.29 | +1.47 | +0.78 |

Supplementary Table 1. Fully relaxed lattice parameters and the distances between different sites (4*e*-4*e* and 4*d*-4*e*) for all the five compounds in comparison to the experimental data. The relaxations are performed in PBE exchange-correlation functional with spin-orbit coupling (SOC).

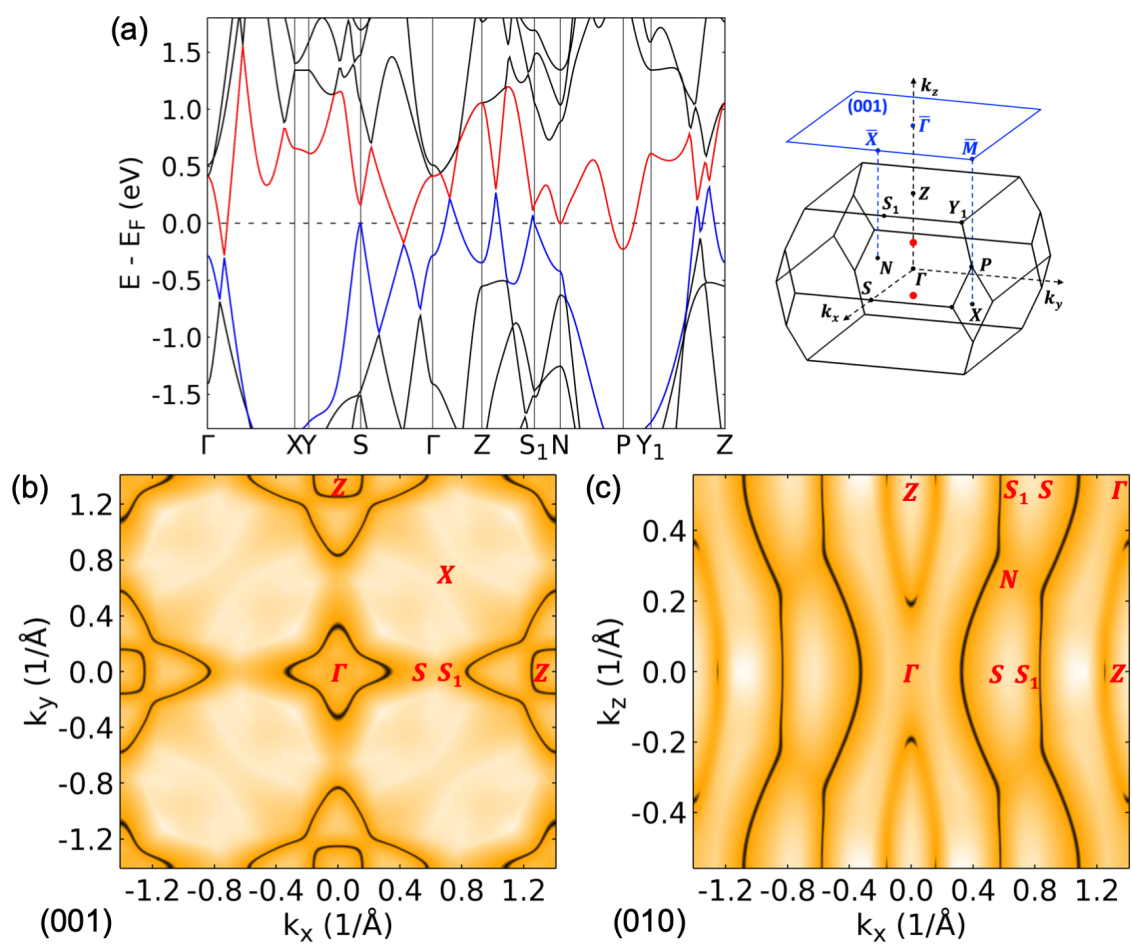

Supplementary Figure 1. Bulk band structure of SrAl$_4$ without SOC and the nodal lines with zero gap on more planes in addition to that in Fig.1(e). (a) Calculated SrAl$_4$ band structure without SOC. The top valence and conduction bands are in blue and red, respectively, with bulk Brillouin zone shown on the right. (b) Nodal lines on the (001) plane shown by the zero gap. (c) Nodal lines on the (010) plane shown by the zero gap.

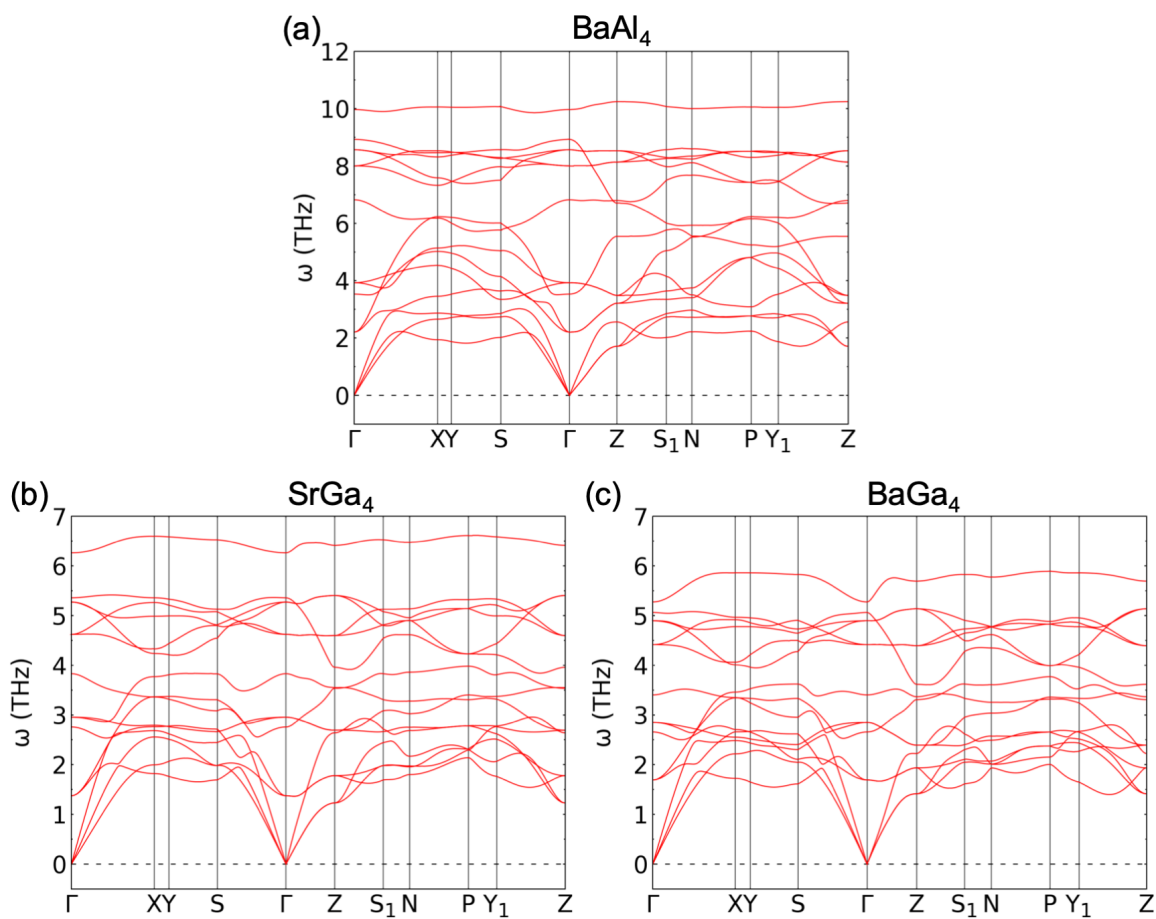

Supplementary Figure 2. Phonon band dispersion of (a) $BaAl_4$, (b) $SrGa_4$ and (c) $BaGa_4$.

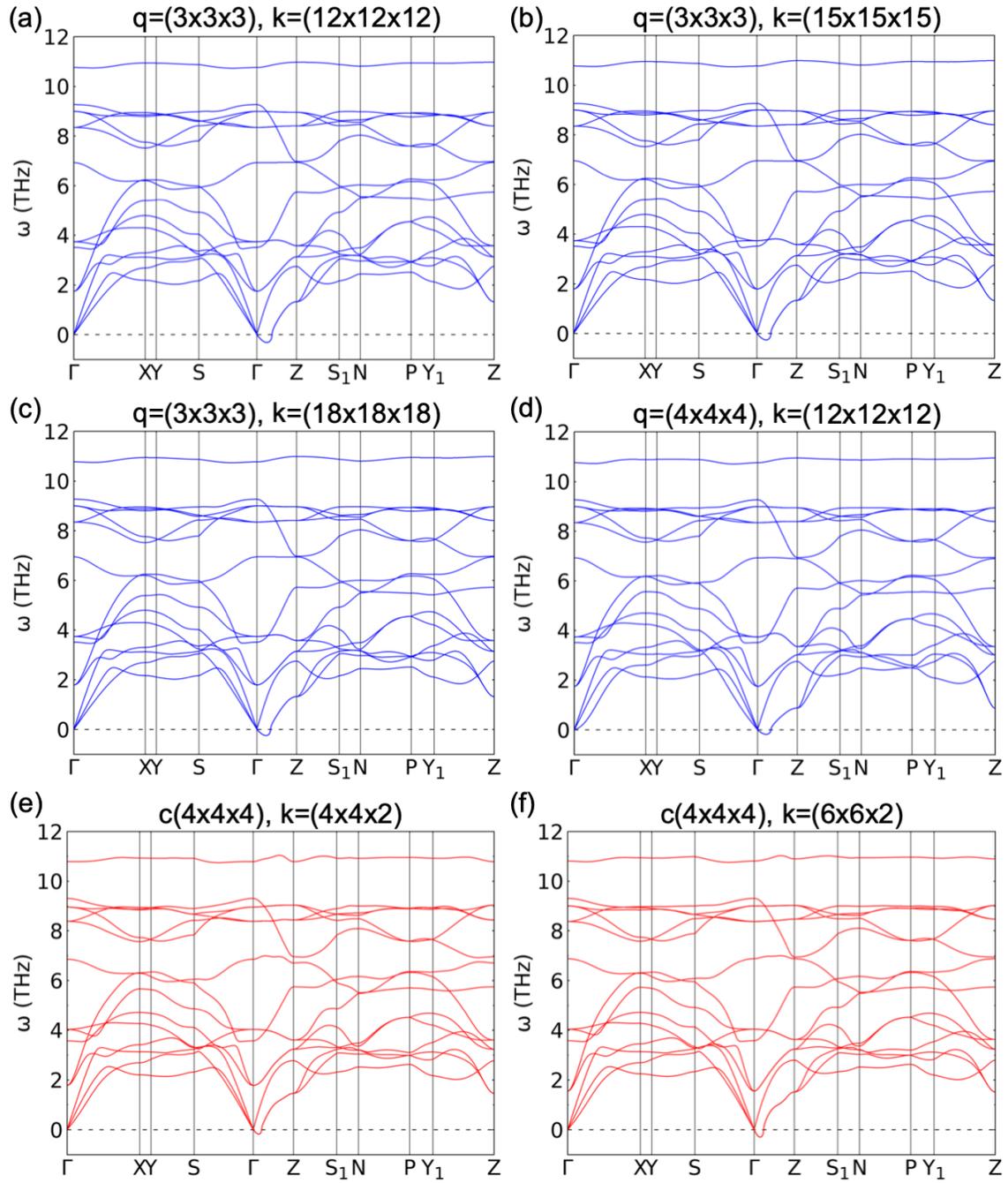

Supplementary Figure 3. The convergence of calculated phonon band dispersions of SrAl$_4$. (a-d) The phonon calculated in QE using DFPT with respect to increasing *k*-mesh and *q*-mesh as labeled. (e-f) The phonon calculated in VASP using the finite displacement method with increasing *k*-mesh as labeled for a large supercell of c(4x4x4) with 640 atoms. All the calculated phonon bands show the imaginary transverse acoustic mode at the small *q*-vector along the Γ-Z direction, as discussed in the main text.

## Supplementary References